\documentclass[twocolumn,aps,showpacs,amsmath,amssymbl]{revtex4}
\usepackage{amsmath,amssymb,bm}
\usepackage{graphicx,color}

\begin{document}

\title{Shear banding of colloidal glasses - a dynamic first order transition?}
\author{V. Chikkadi, D. M. Miedema, B. Nienhuis and P. Schall}

\affiliation{van der Waals- Zeeman Institute, University of Amsterdam, Science Park 904, 1098 XH Amsterdam, The Netherlands\\}

\begin{abstract}
We demonstrate that application of an increasing shear field on a glass leads to an intriguing dynamic first order transition in analogy to equilibrium transitions. By following the particle dynamics as a function of the driving field in a colloidal glass, we identify a critical shear rate upon which the diffusion time scale of the glass exhibits a sudden discontinuity. Using a new dynamic order parameter, we show that this discontinuity is analogous to a first order transition, in which the applied stress acts as the conjugate field on the system's dynamic evolution. These results offer new perspectives to comprehend the generic shear banding instability of a wide range of amorphous materials. \end{abstract}

\pacs{82.70.Dd, 64.70.kj, 62.20.F-, 61.43.Fs}{}

\maketitle

A central unresolved question in the physics of glasses concerns the behavior of a glass under applied stress. While at the glass transition, microscopic observables change rather smoothly, yet rapidly~\cite{Ediger1996,Pusey1987} as a function of density or temperature, an important question is whether a similarly smooth variation occurs upon application of stress. Recent experiments and simulations show that, unlike quiescent glasses, slowly sheared glasses exhibit high dynamic susceptibilities with long-range, directed strain correlations~\cite{Maloney,Caroli,Chikkadi11,Chikkadi12}; such long-range correlations indicate a high susceptibility of the deformation also with respect to the applied shear. The question is then how the glass responds to an increasing applied shear field. Recent work has focused on two main regimes: The thermal regime, where the applied shear rate is slower than the rate of thermally activated rearrangements. This regime exhibits interesting evidence of solidity and elasticity, leading to long-range strain correlations that dominate the flow~\cite{Caroli,Chikkadi11}. Second, the shear-dominated regime, where the applied shear rate is faster than thermal relaxation, and rearrangements are dominated by the applied shear. Here, liquid-like properties are observed: mean-square displacements are described by an effective shear-rate dependent temperature~\cite{Eisenmann2010,besseling10}. This regime is also addressed by many models of plasticity~\cite{STZ} and rheology~\cite{SGR} assuming uncorrelated, liquid-like flow. The two mentioned regimes represent the fundamental solid and liquid properties of glasses. A crucial open question is how these regimes are connected. Investigation of the dynamics of the glass under increasing applied shear rate would allow insight into the transition between the solid and liquid regimes, but remains challenging: in simulations, the limited simulation time scale makes the slow thermally activated steady-state flow difficult to address; in experiments, direct observation of the dynamics in molecular glasses is prohibitively difficult.

Colloidal glasses allow direct observation of single particle dynamics, offering particle trajectories to be followed over long time and large length scales. The constituent particles exhibit dynamic arrest due to crowding at volume fractions larger than $\phi_g \sim 0.58$, the colloidal glass transition~\cite{Pusey1986,vanMegen1998}. These systems exhibit glass-like properties such as non-ergodicity and aging \cite{Bouchaud1992}, and display the characteristic features of solid-like long-range correlations in the thermal~\cite{Chikkadi11}, and liquid-like diffusivity~\cite{Eisenmann2010} in the shear-dominated regime. Experiments performed at increasing shear rates have shown intriguing shear inhomogeneity with symmetry changes of correlations from solid to liquid~\cite{Chikkadi11}. Such symmetry change usually indicates a qualitative change of the material; the crucial open question is if this qualitative change indicates the dynamic analogue of a first order phase transition.

In this letter, we use direct observation of single particle dynamics in a colloidal glass to show that the transition from the thermal to the shear-dominated regime has hallmarks of a dynamic first order transition. We demonstrate the existence of a critical shear rate, at which the glass separates into two dynamic states characterized by distinct diffusion time scales. We measure a new dynamic order parameter~\cite{hedges09} to demonstrate the coexistence of two dynamic phases. Because this order parameter is extensive in space and time, the novelty of this dynamic transition is its occurrence in four-dimensional space-time. We show that this transition is accompanied by weak structural modification of the glass. These results offer a new framework to understand the genuine shear-banding instability observed in a wide range of colloidal and metallic glasses~\cite{schall_hecke10,Greer13}.

The colloidal glass consists of sterically stabilized fluorescent polymethylmethacrylate (PMMA) particles with a diameter of $\sigma = 1.3 \mu m$, and a polydisperity of $ 7 \%$, suspended in a density and refractive index matching mixture of Cycloheptyl Bromide and Cis-Decalin. A dense suspension with particle volume fraction $\phi \sim 0.60$ well inside the glassy state is prepared by diluting suspensions centrifuged to a sediment. The suspension is loaded in a cell between two parallel plates $65 \mu m$ apart, and a piezoelectric translation stage is used to move the top boundary to apply shear at constant rates between $\dot{\gamma} = 1.5 \times 10^{-5}$ and $2.2 \times 10^{-4} s^{-1}$, with a maximum strain of $140\%$. Confocal microscopy is used to image the individual particles, and determine their positions in three dimensions with an accuracy of $0.03\mu m$ in the horizontal, and $0.05\mu m$ in the vertical direction~\cite{weeks_weitz00}. All measurements presented here are recorded in the steady-state regime, after the sample has been sheared to $\gamma \sim 1$. We use the structural relaxation time $\tau = 2 \times 10^4 s$~\cite{Chikkadi11} of the quiescent glass to define the dimensionless shear rate $\dot{\gamma}^* = \dot{\gamma} \tau$; the applied shear rates then correspond to $\dot{\gamma}^*$ between $0.3$ and $2$, smaller and larger than one.

%%%%%%%%%%%%%%%%%%%%%%%%%%%%%%%%%%%%
\begin{figure}
\centering
\includegraphics[width=0.5\textwidth]{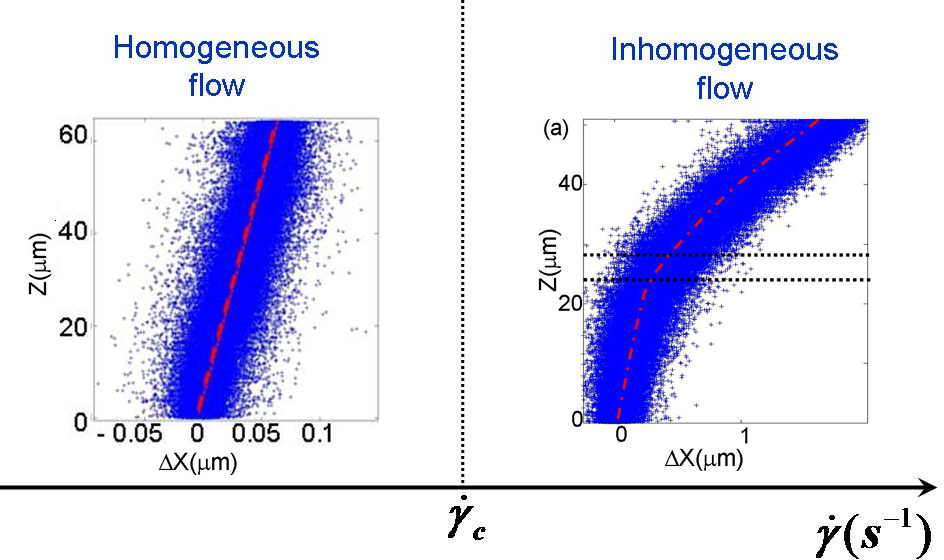}
\caption{Deformation map of colloidal glasses at a volume fraction $\phi=0.60$. The flow is homogeneous at low shear rates (left), and inhomogeneous beyond the critical shear rate $\dot{\gamma}_c\sim 6\times10^{-5}s^{-1}$ (right). The figures show the height-dependent displacements of the particles at the shear rates $\dot\gamma = 3 \times 10^{-5}s^{-1}$ (left) and $\dot\gamma = 1 \times 10^{-4}s^{-1}$ (right). Each cross represents a particle. Dashed horizontal lines (right) define the boundaries of the low and high shear bands, and their interface region.}
\label{fig1}
\end{figure}
%%%%%%%%%%%%%%%%%%%%%%%%%%%%%%%%%%%%%

The shear-rate dependent flow behavior is summarized in Fig.~\ref{fig1}. At shear rates $\dot{\gamma}^* < 1$, the glass flows homogeneously as shown by the particle displacements as a function of height in Fig. 1a. At $\dot{\gamma}^* > 1$, the glass separates spontaneously into bands that flow at different rates as shown for $\dot{\gamma}^* = 2$ in Fig.~\ref{fig1}b. Linear fits to the displacement profiles yield flow rates of $\dot\gamma_{\text{high}} = 2.2 \times 10^{-4}s^{-1}$ and $\dot\gamma_{\text{low}} = 4 \times 10^{-5}s^{-1}$ that differ by a factor of 5. By investigating the flow profile over the entire observation time, we confirm that the slopes remain unchanged after an initial transient, indicating steady state coexistence. We thus observe the spontaneous transition from steady-state homogeneous to steady-state inhomogeneous flow at $\dot{\gamma}^* \sim 1$. This transition from homogeneous to inhomogeneous flow is analogous to the shear banding in metallic glasses~\cite{spaepen75,spaepen77}.

\begin{figure}
\begin{center}
\includegraphics[width=0.45\textwidth]{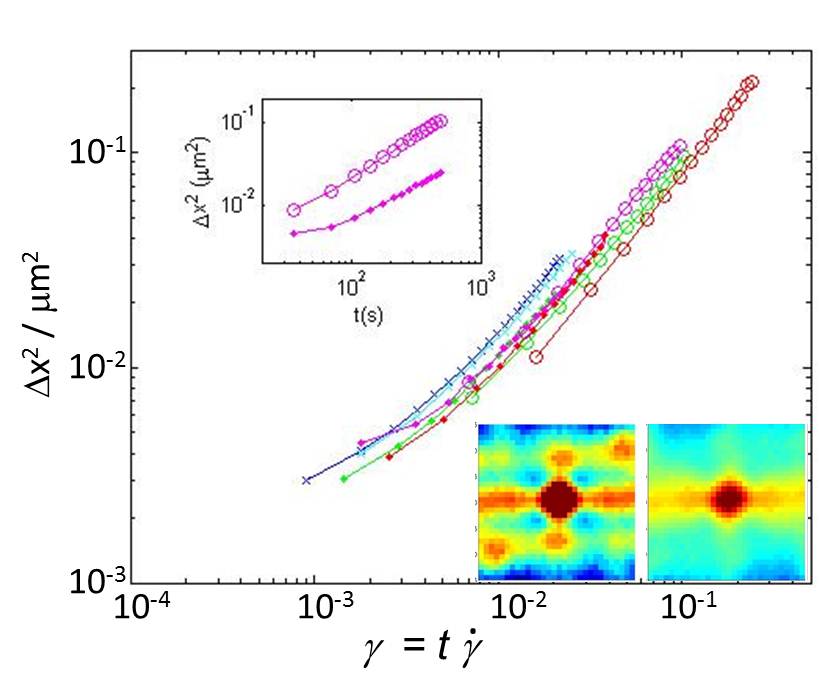}
\end{center}
\caption{Mean square displacements of the particles. Upper left inset: mean square displacement in the upper (circles) and lower shear band (dots) at $\dot{\gamma}^* = 2$. Main panel: mean-square displacements as a function of rescaled time for the applied shear rates $\dot{\gamma}^* = 0.3$ (blue), 0.6 (cyan), 1.2 (green), 2 (magenta) and 5.6 (red). Lower right inset: strain correlations in the low shear band calculated for $t = 70~s$ (left) and $350~s$ (right).}
\label{fig2}
\end{figure}

To elucidate it, we take advantage of the recorded particle trajectories and investigate the particle dynamics as a function of the applied shear. We use the trajectory ${\bf \Delta r_i}(t)$ of each particle to compute its displacement fluctuations around the mean flow, ${\bf \Delta r'_{i}}(t) = {\bf \Delta r_i}(t) - \langle{\bf \Delta r}(t)\rangle_z$, where $\langle{\bf \Delta r}(t)\rangle_z$ is the average particle displacement at height $z$. Typical examples of the resulting mean-square displacements $\langle{\bf \Delta r'}^2(t)\rangle$ in the high and low shear band are shown in Fig.~\ref{fig2} (inset). The low shear band (stars) reveals reminiscence of a plateau, while the high shear band (circles) exhibits a closely linear increase of $\langle{\bf \Delta r'}^2(t)\rangle$ similar to the mean-square displacement of particles in a liquid. Strain correlations computed separately for the two bands show coexistence of an isotropic liquid-like to an anisotropic solid-like response~\cite{Chikkadi11}. Similar behavior is observed for all other shear rates with $\dot{\gamma}^* > 1$. Interestingly, we can collapse all mean-square displacements by re-scaling the time axis by $\dot{\gamma}$ as shown in Fig.~\ref{fig2}, main panel. The figure compiles measurements both in the homogeneous and shear-banding regime. This collapse suggests that the different dynamics of the bands is solely due to different underlying diffusion time scales; indeed, this is supported by the strain correlation function of the low shear band shown in Fig.~\ref{fig2} (lower right insets): the solid-like quadrupolar symmetry disappears when correlations are computed on the rescaled time scale (longer by a factor of $\dot\gamma_{\text{high}} / \dot\gamma_{\text{low}}$). We thus conclude that the change of diffusion time scale causes the symmetry change of correlations; such discontinuous change reminds of first order transitions, with the discontinuity occurring in the underlying diffusion time scale.
%%%%%%%%%%%%%%%%%%%%%%%%%%%%%
%%%%%%%%%%%%%%%%%%%%%%%%%%%%%

\begin{figure}
\centering
\includegraphics[width=0.47\textwidth]{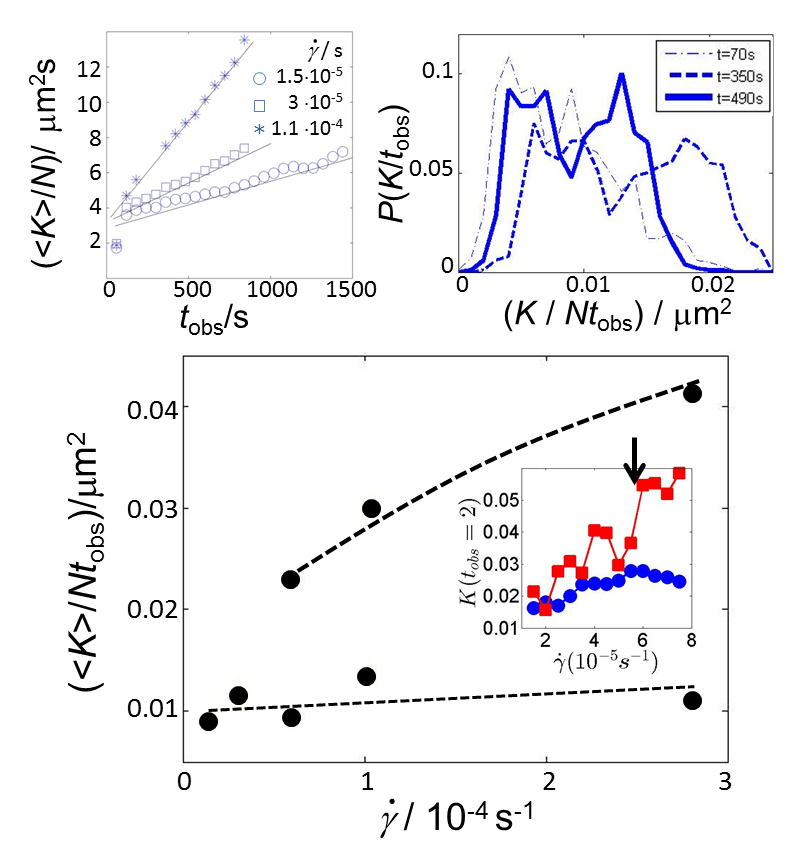}
\begin{picture}(0,0)(0,0)
\put(-234,155){(a)}
\put(-120,155){(b)}
\put(-234,10){(c)}
\end{picture}
\caption{Dynamic order parameter and phase diagram. (a) Dynamic order parameter as a function of observation time. The linear relations demonstrate that $K$ is a linear measure of the system's dynamic evolution. (b) Histogram of order parameter values for increasing observation times. The emerging bimodal distribution indicates dynamic phase coexistence. (c) Corresponding dynamic phase diagram: Mean order parameter as a function of applied strain rate. The dashed lines delineate boundaries of the shear banding regime. Inset shows the dynamic order parameter as a function of time at continuously increasing applied shear rate, for particles in the upper (red squares) and lower region (blue dots). Arrow demarcates a sudden change of the order parameter.}
\label{fig3}
\end{figure}
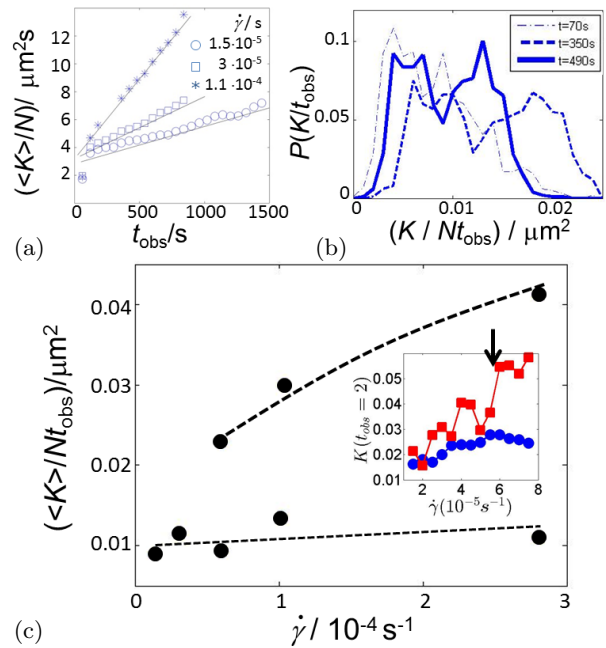

To quantify this dynamic discontinuity, we search for an order parameter that is a good measure of the dynamic evolution. An appropriate measure of the underlying dynamic evolution is ~\cite{hedges09}
%%%%%%%%%%%%%%%%%%%%%%%%%%%%%%%
\begin{equation}
\label{OP_K} K = \Delta t {\sum_{i=1}^{N}} {\sum_{t=0}^{t_{obs}}} \vert
{\bf \Delta r_i'}(t+  \Delta t) - {\bf \Delta r_i'}(t) \vert^2,
\end{equation}
%%%%%%%%%%%%%%%%%%%%%%%%%%%%%%%%
the time-integrated mean-square displacement, where $\Delta t$ is a short microscopic time scale. This order parameter is extensive in time, i.e. increases linearly with observation time $t_{obs}$, as shown in Fig.~\ref{fig3}a.  Thus, $K/t_{obs}$ measures the rate of the system's dynamic evolution. To address the transition, we determine values of $K$ in 2$\mu$m thick horizontal subsections, and plot probability distributions of $K/t_{obs}$ for three different observation times in Fig.~\ref{fig3}b. With increasing observation time, two peaks appear and sharpen, demonstrating the coexistence of two dynamic states. The peak positions demarcate the order parameter values of the coexisting shear bands and reveal their different dynamic evolution. We can now construct the corresponding dynamic phase diagram: by determining the peak values of $K$ for all steady-state shear rates, we obtain the corresponding diagram as shown in Fig.~\ref{fig3}(c). At $\dot{\gamma}^* < 1$, only one single peak of $K$ exists, indicating the homogeneous regime. At $\dot{\gamma}^* > 1$, two values coexist, indicating the coexisting shear bands. The diagram has the characteristic topology of a phase diagram, in which the two-phase region is entered close to a critical point. A related dynamic phase coexistence has been recently observed by us in traffic models with interacting cars~\cite{deWijn}. With increasing density and in the limit of strong braking, traffic jams exhibit long-range correlations, after which macroscopic phase separation into jammed and free moving traffic occurred.

The data in Fig. \ref{fig3}c does not allow us to see clearly the transition from the single phase to the two-phase region; this would require closely spaced steady-state measurements. To elucidate the sharpness of the dynamic transition as a function of the applied field, we therefore continuously increased the shear rate $\dot{\gamma^*}$ from below to above 1 in a single experiment, crossing the transition with a continuously increasing shear field. The resulting values of $K$ as a function of strain rate (Fig.~\ref{fig3}c, inset) indicate that indeed, the transition occurs rapidly. Because of the increasing strain rate, we can only use short observation time intervals, $t_{obs} = 2$, leading to significant fluctuations of $K$; nevertheless, the data suggests a sudden jump of the order parameter, indicating an abrupt transition at $\dot{\gamma^*} \sim 1$. A related sharp transition was recently observed by us in the oscillatory yielding of colloidal glasses; the structure factor exhibited an abrupt symmetry change from anisotropic solid to isotropic liquid when strain amplitudes exceeded the yielding threshold~\cite{Denisov13}.

%%%%%%%%%%%%%%%%%%%%%%%%%
%%%%%%%%%%%%%%%%%%%%%%%%

This sharp dynamic transition is surprising and suggests highly collective dynamic behavior of the system. Dynamic first-order transitions were recently observed in simulations on facilitated glassy dynamics~\cite{hedges09}: under an applied artificial field $s$ that couples to the dynamic order parameter distribution via $P(s)=P_0 \exp(-Ks/kT)$, where $P_0$ is the unperturbed distribution, all hallmarks of a true first order transition were observed. In the present case, the applied shear stress $\sigma$ can take the role of the conjugate field: it is well known that the applied stress $\sigma$ couples to local rearrangements via their induced strain $\epsilon$ according to $P(\epsilon)=P_0 \exp(-\sigma \Omega(\epsilon)/kT)$, where the activation volume $\Omega(\epsilon) = \int \epsilon dV$ measures  the integrated local strain~\cite{Eyring,spaepen77}. Because of the high dynamic susceptibility of the colloidal glass under applied shear~\cite{Chikkadi11}, likewise, the coupling between the applied stress and the particle dynamics can introduce a first-order transition in its dynamic evolution.

%accumulated strain is related to $K$, we conclude that the application of a mechanical field on a system with facilitated dynamics can provide the sharp dynamic transition observed here.

%%%%%%%%%%%%%%%%%%%%%%%%%%%%%%%%%%%%
\begin{figure}
\centering
\begin{minipage}{5.0cm}
\includegraphics[width=1\textwidth]{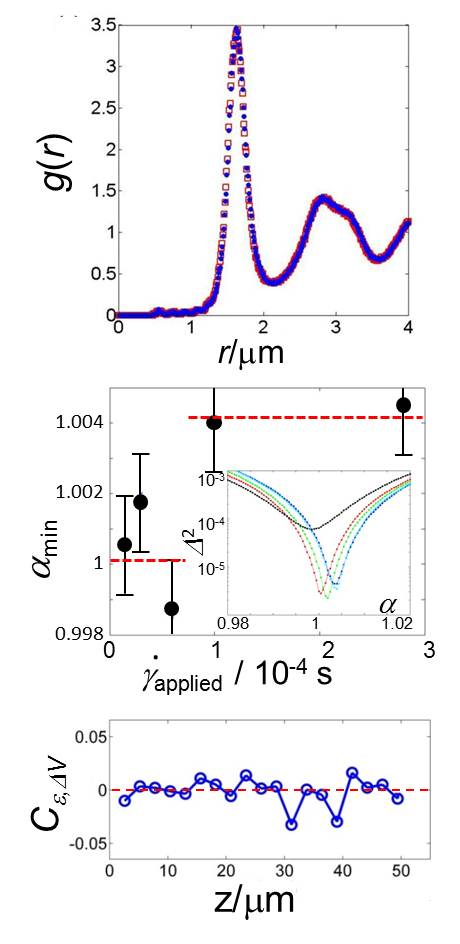}
\put(-132,176){(a)}
\put(-132,76){(b)}
\put(-132,9){(c)}
\end{minipage}
\caption{Glass structure and density. (a) Pair correlation function of particles in the low (blue dots) and high shear band (red squares). The close overlap demonstrates the robustness of the sheared glass structure. (b) Dilation parameter, $\alpha_{min}$, as a function of applied shear rate determined from the minimum of the mean-square difference of the $g(r)$ curves as shown in the inset. Dashed lines are guides to the eye. (c) Normalized correlation between local shear and dilation averaged over all particles.}
\label{fig4}
\end{figure}
%%%%%%%%%%%%%%%%%%%%%%%%%%%%%%%%%%%%%

While the transition occurs in the dynamics, it is interesting to elucidate changes in the glass structure. Constitutive models of the flow of amorphous materials suggest a coupling between flow and structure, often related to small density changes~\cite{spaepen77,STZ,besseling10}; we therefore investigated structural differences in the two bands. We show radial distribution functions in Fig.~\ref{fig4}a. No obvious structural difference between the low and high shear band is observed, in agreement with earlier observations~\cite{besseling10}. However, to check for small amounts of dilation, we compute the mean-square difference $\Delta^2$ between the two $g(r)$ curves as a function of a linear stretching $\alpha$ that transforms $r$ to $r' = r \times \alpha$. We show $\Delta^2$ as a function of $\alpha$ in Fig.~\ref{fig4}b, inset; the minima of $\Delta^2$ at $\alpha > 1$ indicate a small amount of dilation. We evaluate the minima $\alpha_{min}$ for all shear rates and plot $\alpha_{min}$ as a function of shear rate in the main panel. These values indicate a dilation of $\sim 0.4\%$ in the high-shear band after shear banding. While the detected changes are small and affected by large uncertainty, they demonstrate a structural change accompanying the shear banding transition. Yet, this structural coupling does not transfer to local quantities. To check directly for a local coupling between shear and dilation, we computed the local strain tensor ${\bf \epsilon}$ for each particle from the change of nearest neighbor vectors over time~\cite{falk_langer98,schall07}. We then computed correlations between the shear and dilation components of the strain tensors for each particle. The resultant average correlation coefficient depicted in Fig.~\ref{fig4}c does not show any significant correlation, thereby indicating -- within the experimental accuracy -- no direct coupling between local shear and dilation.
%in agreement with recent computer simulations on sheared glasses~\cite{varnik2012}.

The direct observation of particle dynamics during shear banding of a colloidal glass suggests that shear banding can be interpreted as a dynamic first order transition. The applied shear plays the role of a conjugate field that couples to the dynamic evolution.
% causing a discontinuous change in the diffusion time scale.
Sufficiently high values of the applied shear rate cause coexistence of two regimes with different time scales for diffusion. This mechanism points out new perspectives to comprehend flow instabilities in amorphous materials: the large  dynamic susceptibility on the one hand (evidenced by long-range strain correlations), and the coupling to the applied shear on the other hand lead to a dynamic transition that is akin to first order transitions. We believe that the presented dynamic first order transition should be a general feature of dynamically driven systems, and recent traffic simulations reveal the formation of dynamic condensates in one-dimensional dynamically facilitated systems, consistent with this idea~\cite{deWijn}. The observed coupling between applied stress and diffusion time could play a role in crowded biological systems; as the diffusion time scale is an important underlying time scale, any direct coupling of diffusion to external (shear) fields would greatly affect the diffusive behavior upon mechanical perturbation.

\end{document}